\documentclass{ifacconf}
\usepackage{graphicx}
\usepackage{amsmath}
\usepackage{amssymb}
\usepackage{natbib}        
\usepackage{siunitx}

\begin{document}

\begin{frontmatter}

\title{Cross-Directional Modelling and Control of Slot-Die Battery Electrode Coating}

\thanks[footnoteinfo]{This work was supported by the Faraday Institution NEXTRODE project [grant number: FIRG090].}

\author[ox]{Hyuntae Kim} and 
\author[ox]{Idris Kempf}
\address[ox]{Department of Engineering Science, University of Oxford, Oxford OX1 3PJ, UK. Email: \{hyuntae.kim, idris.kempf\}@eng.ox.ac.uk}

\begin{abstract}
As global battery demand increases, real-time process control becomes increasingly important for battery electrode manufacturing, yet slot-die lines are still mostly manually operated in open loop. This paper develops a physics-based modelling-and-control pipeline for film-thickness regulation. Computational fluid dynamics (CFD) simulations provide the data from which a low-order cross-directional model is identified and calibrated. Numerical simulations demonstrate close agreement between the CFD and the cross-directional model, which is used to design a controller that can be used in both real-time, automated feedback operation and manual feedforward operation during line commissioning.
\end{abstract}


\end{frontmatter}

\section{Introduction}

Electrode manufacturing is a major lever for improving the cost, carbon footprint, and material efficiency of lithium-ion batteries \citep{Grant2022,Hawley2019,Li2011,Tarascon2001,Arora1998}. However, many industrial lines are still operated in open loop: coating, drying, and calendering recipes are tuned offline, with limited feedback. As global battery demand increases, recent work argues for optimised processes, in which thickness, loading, and porosity are regulated directly along the line \citep{Hallemans2025}. This is particularly important for slot-die coating, where film nonuniformity and drift translate directly into scrap and lost capacity \citep{Kistler1997,Ruschak1985,Schmitt2013a,Schmitt2013b,Gong2024,Kasischke2021}.

Realising such product-centric control faces both practical and modelling challenges. Practically, electrode lines provide sparse in-line metrology, and the cost of additional sensing and automation must first be justified economically~\citep{Reynolds2021}. Technically, high-fidelity multiphase CFD can resolve bead dynamics and operating windows in detail \citep{Kistler1997,Ruschak1985,Schmitt2013a,Schmitt2013b,Gong2024,Kasischke2021}, but is mathematically and computationally too expensive for real-time control in electrode plants \citep{Grant2022,Hawley2019}. Empirical low-order models are easy to embed in controllers but often obscure the underlying physics and extrapolate poorly beyond the identification regime \citep{Hawley2019,Li2011}. There is therefore a need for compact, PDE-informed models that retain the essential structure of the governing equations, can be adapted to different slot-die geometries, and remain transparent enough for feedback design using standard control tools.

This paper takes a step toward PDE-informed modelling and feedback control for slot-die coating in lithium-ion electrode manufacturing, by deriving a low-order convective–relaxation model tied to the slot-die geometry and using it for cross-directional thickness regulation. We focus on slot-die coaters equipped with several independently metered feed zones: the inlet flow is adjusted in multiple stripes across the coating width, a configuration that can be extended to accommodate a flex-lip slot die. Starting from a two-phase incompressible Navier-Stokes \textit{volume of fluid} (VOF) model of a slot-die coating slice \citep{Kistler1997,Ruschak1985}, we derive a low-order surrogate consisting of a convective transport delay, second-order convective-relaxation dynamics, and a static cross-directional DC gain matrix, with parameters calibrated from the same CFD model. We demonstrate in numerical simulations that the output of the cross-directional approximation closely matches the one of the CFD model. We then design a proportional (P) controller based on the cross-directional DC gain matrix, with a single scalar tuning parameter that scales a decoupling-type gain. Closed-loop simulations show that the resulting feedback achieves well-damped tracking of thickness set-points with a nearly uniform cross-directional profile, consistent with the product-control objectives advocated in recent battery manufacturing studies \citep{Hallemans2025,Grant2022,Hawley2019,Li2011,Tarascon2001,Arora1998}. In plants where real-time monitoring and/or automated actuation is not implemented yet, the cross-directional model and controller can be used to manually configure the multi-input slot-die.

This paper is structured as follows. The PDEs of a standard Navier-Stokes-VOF model are summarised in Section~\ref{sec:pde-model}, and are used in Section~\ref{sec:pde-to-ode} to derive a lower-order, cross-directional model of the coating thickness. Section~\ref{sec:pde-informed-H} calibrates the scalar dynamics and the cross-directional gain matrix on an example slot-die geometry, and Section~\ref{sec:time-domain-validation} validates the surrogate against the CFD data and illustrates a simple feedback design.

\section{Navier-Stokes Equations}\label{sec:pde-model}

\begin{figure}
    \centering
    \includegraphics[width=0.9\linewidth]{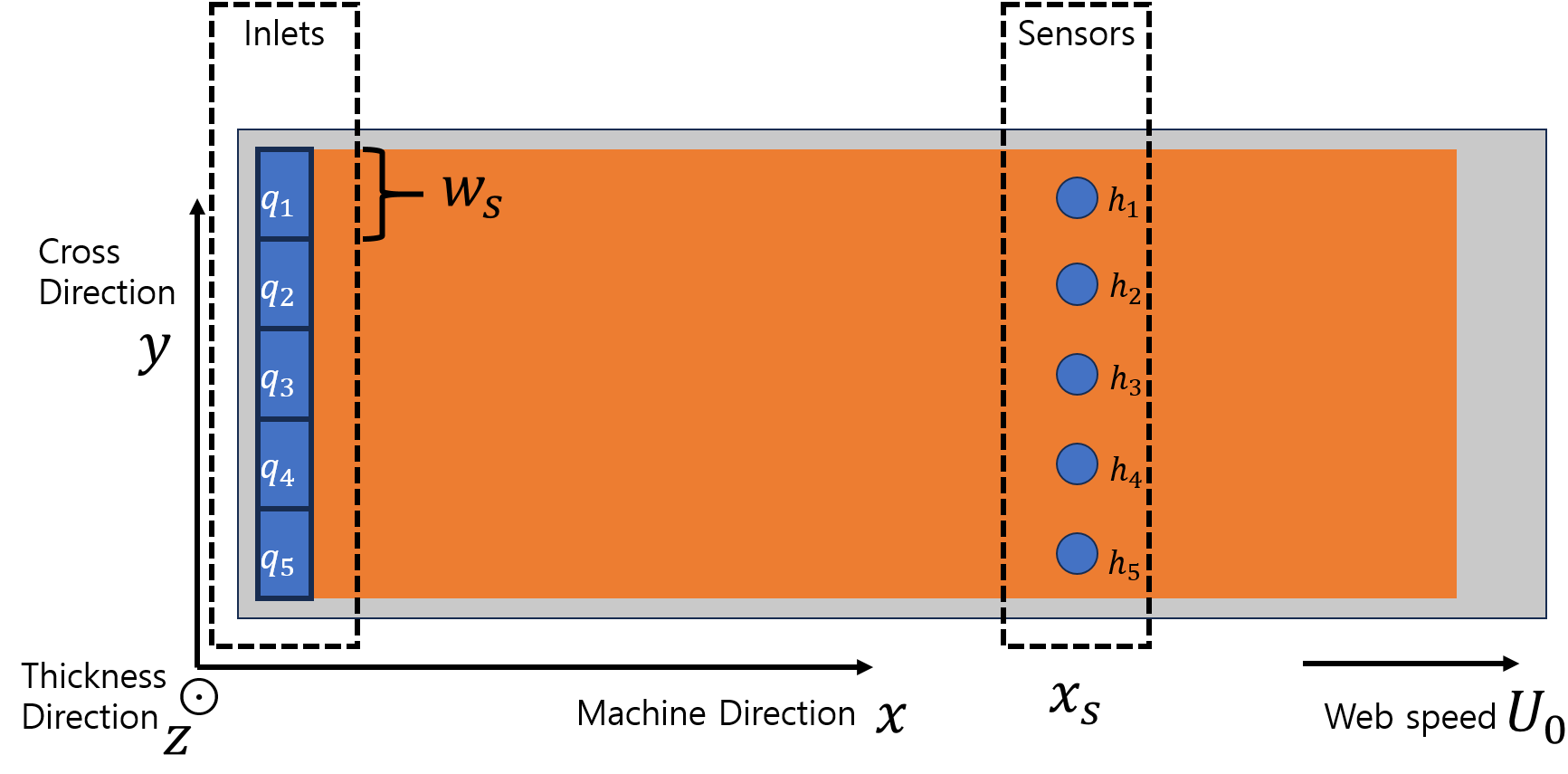}
    \caption{Schematics for slot-die coating model.}
    \label{fig:line}
\end{figure}

To describe how the wet-film thickness is affected by coating process variables, we work with a physics-based flow model governed by PDEs. The coating is modelled as two immiscible, incompressible phases (liquid and air) in a fixed three-dimensional coordinate system, $\mathbf{x} = (x,y,z)^{\top}$, where $x$ is the machine direction and $y$ is the cross direction. A simplified layout of the slot-die is shown in Fig.~\ref{fig:line}. The wet-film thickness, $\mathbf{h}(t) := [\,h_1(t),\dots,h_n(t)\,]^{\top}\in\mathbb{R}^n$ with $h_j(t) := h(x_s,y_j,t)$, is measured by $n$ sensors located at $x_s$ and $y_j$. The coating thickness results from the interplay of acceleration under pressure differences, viscosity, gravity, and surface tensions with the web speed $U_0$ and the $m$ coating inlet flows, $\mathbf{q}(t)=\big[q_1(t),\dots,q_m(t)\big]^{\top}$, where $q_i(t)$ in \si{\meter^3\per\second} is modelled as the total liquid flow through the $i$-th inlet stripe of width $w_s$. In the following it is assumed that there are as many sensors as inlets ($m=n$) and that the sensors are aligned with the inlets as in Fig.~\ref{fig:line}, but the analysis remains the same for $m\neq n$ and different sensor and inlet geometries.

The phase distribution is represented by a volume-of-fluid (VOF) indicator $\alpha(\mathbf{x},t)\in[0,1]$, where $\alpha=1$ in the liquid and $\alpha=0$ in the air. The primary fields are the mixture velocity $\mathbf{u}(\mathbf{x},t)\in\mathbb{R}^3$ and pressure $p(\mathbf{x},t)\in\mathbb{R}$. These are assumed to obey the incompressible Navier-Stokes equations
\begin{align*}
\rho(\alpha)\Big(\frac{\partial \mathbf{u}}{\partial t}+\mathbf{u}\cdot\nabla \mathbf{u}\Big)
&= -\nabla p + \nabla\!\cdot\!\big(\mu(\alpha)\,[\nabla \mathbf{u}+\nabla \mathbf{u}^{\top}]\big)\nonumber\\
&\quad\ -\rho(\alpha)\,g\,\mathbf{e}_z+\mathbf{f}_{\sigma}, \nonumber\\
\nabla\cdot \mathbf{u} &= 0,
\end{align*}
where $\mathbf{f}_{\sigma}$ is the capillary body force that models surface-tension effects at the liquid-air interface and the incompressibility condition $\nabla\cdot\mathbf{u}=0$ enforces conservation of volume for the mixture. The inlet flows define a boundary condition at $x=0$. The mixture density and viscosity are obtained from the local phase fraction,
\[
\rho(\alpha)=\alpha\,\rho_{\ell}+(1-\alpha)\,\rho_{a},\qquad
\mu(\alpha)=\alpha\,\mu_{\ell}+(1-\alpha)\,\mu_{a},
\]
where $\rho_{\ell}$ and $\mu_{\ell}$ denote the density and viscosity of the liquid phase, and $\rho_{a}$ and $\mu_{a}$ those of the air phase, so that liquid-rich regions are heavier and more
viscous than air-rich regions. 

The Navier-Stokes VOF model is implemented and solved using the \texttt{interFoam} solver from the OpenFOAM library \citep{Weller1998}. The computational domain $\Omega\subset\mathbb{R}^3$ is a thin slice of the slot-die geometry, with vertical extent $0\le z\le H_z$. The primary fields are the mixture velocity $\mathbf{u}(\mathbf{x},t)\in\mathbb{R}^3$ and the pressure $p(\mathbf{x},t)\in\mathbb{R}$, and the phase distribution is represented by the VOF indicator $\alpha(\mathbf{x},t)\in[0,1]$. The indicator $\alpha$ is advected with the same mixture velocity field $\mathbf{u}(\mathbf{x},t)$, using the conservative advection and interface-compression schemes built into \texttt{interFoam}, so that the liquid–air interface remains sharp and satisfies the prescribed contact angles at solid walls. In the reduced, depth-averaged model introduced later, we retain this convective structure but replace the shear-rate-dependent Carreau–Yasuda liquid rheology by an effective constant viscosity at the nominal operating point, so that the surrogate dynamics remain tied to the full Navier–Stokes–VOF system described here.

The geometry is a thin ``2.5D'' slice aligned with the slot-die and moving web. The substrate at $z=0$ is a rigid translating no-slip wall moving with constant speed $U_0$ in the $+x$-direction, the free surface at $z=H_z$ is open to the atmosphere (pressure outlet), faces normal to $y$ are symmetry planes, and inlet and outlet boundaries lie on the $x$-faces. Following the VOF-based post-processing used in the simulations, we introduce 
a depth-averaged thickness field
\begin{equation}
h(x,y,t)
:= \frac{1}{A_{\mathrm{sens}}}\int_{\mathcal{S}(x,y)}\alpha(\mathbf{x},t)\,dV,
\label{eq:thickness_field}
\end{equation}
where $\mathcal{S}(x,y)\subset\Omega$ is a right cylinder of fixed cross-sectional 
area $A_{\mathrm{sens}}=\pi r_c^2$ centred at $(x,y)$ and oriented in the $z$-direction, 
and $r_c$ is the sensor radius. When the interface is locally flat in $z$, the mapping \eqref{eq:thickness_field} 
(which is a functional that takes the VOF field $\alpha$ and returns a scalar thickness) coincides with the geometric wet-film thickness under the footprint of 
$\mathcal{S}(x,y)$, so $h(x,y,t)$ can be viewed as the local depth-averaged film 
thickness and $h_j(t)$ as its sample at $(x_s,y_j)$.

\section{Cross-Directional Model}
\label{sec:pde-to-ode}

Starting from the Navier-Stokes-VOF model with inlet flow vector $\mathbf{q}(t)$ and thickness measurements $\mathbf{h}(t)$, we seek a low-order surrogate that captures the inlet-to-sensor dynamics along the machine direction and the cross-directional coupling, i.e. the effect of $q_i$ on $h_j$. Because the coating layer is extremely thin compared with the in-plane length scales, variations in the thickness direction $z$ relax on a time scale that is much shorter than the convective transport time along $x$, so we depth-average over $z$ and describe the film by an in-plane thickness field $h(x,y,t)$. To obtain a tractable surrogate, we separate streamwise and cross-directional effects: for a representative inlet-sensor pair at fixed $y=y_i$, we approximate the inlet-to-sensor dynamics along the machine direction by a one-dimensional surrogate $h(x,t) := h(x,y_i,t)$, yielding a scalar transfer function $G(s)$ from $q(t)$ to $h(x_s,y_i,t)$, and we represent the cross-directional influence of the inlet flows on the sensor thicknesses $h_j(t)=h(x_s,y_j,t)$ through a static cross-directional DC gain matrix $\mathbf{H}$ that collects the steady-state sensitivities $\partial h_j/\partial q_i$.

\subsection{Machine direction dynamics}

To obtain a one-dimensional surrogate along the machine direction, we focus on a single inlet stripe with thickness profile $h(x,t) := h(x,y_i,t)$. Because the hardware layout and operating conditions are approximately uniform across stripes, the scalar inlet-to-sensor dynamics are the same for each channel. A standard depth-averaged mass balance for thin films (see, e.g., \citep{Kistler1997,Ruschak1985}) establishes the continuity relation
\begin{equation}
\frac{\partial h}{\partial t} + \frac{\partial J}{\partial x} = 0,
\label{eq:thinfilm_mass_balance}
\end{equation} 
where the per-unit-width volumetric flow rate $J(x,t)$ (in \si{\meter^2\per\second}) is used instead of the total inlet flows $q_i(t)$ (in \si{\meter^3\per\second}).

In slot-die coating, the per-width flow can be decomposed into a dominant
convective part transported by the web moving at speed $U_0$ and a smaller contribution due to levelling,
\begin{equation}
J(x,t) = U_0\,h(x,t) + J_{\mathrm{lev}}(h,\partial h/\partial x,\dots),
\label{eq:J_decomposition}
\end{equation}
where $J_{\mathrm{lev}}$ collects the effects of
gravity, viscosity, and surface tension. Substituting~\eqref{eq:J_decomposition}
into~\eqref{eq:thinfilm_mass_balance} yields the convective-relaxation model
\begin{equation}
\frac{\partial h}{\partial t} + U_0 \frac{\partial h}{\partial x}
= \mathcal{L}_{\mathrm{lev}}[h],
\label{eq:convective_relaxation_PDE}
\end{equation}
where $\mathcal{L}_{\mathrm{lev}}[h]
:= -\frac{\partial}{\partial x}\, J_{\mathrm{lev}}(h,\partial h/\partial x,\dots)$ defines a spatial operator.

To connect~\eqref{eq:convective_relaxation_PDE} to the inlet flow and the downstream sensor, we restrict $x$ to $x\in[0,x_s]$, with $x=0$ and $x=x_s$ representing the die lip and sensor location, respectively. Let $q(t)$ denote the total volumetric inlet flow of the stripe, and let $w_s>0$ denote the inlet width in the cross direction. The inlet flow imposes a boundary condition at $x=0$,
\begin{equation}
J(0,t) = \frac{1}{w_s}\,q(t),
\label{eq:flux_BC}
\end{equation}
and the physical sensor output is the local thickness $h(x_s,t)$ at $x=x_s$. To linearise \eqref{eq:convective_relaxation_PDE}-\eqref{eq:flux_BC} about the steady film $h_0$ and flow $q_0$, we write $q(t) = q_0 + \delta q(t)$ and $\tilde h(x,t)=h(x,t)-h_0$. We work in deviation coordinates, so the output of the linearised model is the thickness perturbation at the sensor,
\[
y(t) := \tilde h(x_s,t),
\]
while the physical thickness is $h(x_s,t) = h_0 + \tilde h(x_s,t)$. In these coordinates the linearised dynamics can be written schematically as
\begin{equation}
\begin{aligned}
\frac{d}{dt}\tilde h(\cdot,t) &= \mathcal{A}\,\tilde h(\cdot,t) + \mathcal{B}\,\delta q(t),\\
y(t) &= \mathcal{C}\,\tilde h(\cdot,t),
\end{aligned}
\label{eq:abstract_state_space}
\end{equation}
where $\mathcal{A}$ is the operator associated with the convective-relaxation PDE \eqref{eq:convective_relaxation_PDE}, $\mathcal{B}$ encodes the boundary input \eqref{eq:flux_BC}, and $\mathcal{C}$ evaluates the thickness perturbation at $x=x_s$; see, for example, \citep{Curtain1995} for a semigroup formulation of such thin-film flow models. In the absence of levelling ($\mathcal{L}_{\mathrm{lev}}=0$) the solution is transported downstream with speed $U_0$, so a perturbation at $x=0$ reaches the sensor at $x=x_s$ after the convective transport delay $L = x_s/U_0$.

We now approximate the delay-free part of this linear PDE system by a
finite-dimensional model. Guided by the hardware layout and the CFD results, we
retain a small number of dominant storage modes: effective compressible storage
in the hardware that feeds the die and storage
of liquid volume in the near-die region where the liquid is guided onto the
moving substrate. A Galerkin projection of the PDE and boundary conditions onto
these dominant storage modes yields a two-state delay-free approximation (see,
e.g., \citep{Kistler1997,Ruschak1985,Curtain1995})
\begin{equation}
\begin{aligned}
\dot \zeta(t) &= A \zeta(t) + B\,\delta q(t),\\
\delta h(x_s,t) &= C \zeta(t),
\end{aligned}
\label{eq:two_state_model}
\end{equation}
where
\[
\delta h(x_s,t) := \tilde h(x_s,t) = h(x_s,t) - h_0
\]
denotes the thickness perturbation at the sensor in deviation coordinates.
The associated delay-free transfer function from $\delta q$ to $\delta h(x_s,\cdot)$ is
\begin{equation}
G_{0}(s)
= C(sI-A)^{-1}B
= \frac{b_0 + b_1 s}{s^2 + c_1 s + c_0}.
\label{eq:G_relax_rational}
\end{equation}
The denominator coefficients $c_0$ and $c_1$ can be interpreted as effective
stiffness and damping of the coupled coating-supply system described by
\eqref{eq:convective_relaxation_PDE}-\eqref{eq:two_state_model}, while the
numerator coefficients $b_0$ and $b_1$ describe how the inlet flow excites
these modes and how they are combined in the measured thickness. Combining the convective transport delay with~\eqref{eq:G_relax_rational} yields the inlet-to-sensor dynamics along the machine direction:
\begin{equation}
G(s)\;:=\;e^{-L s}\,G_{0}(s).
\label{eq:G_def}
\end{equation}

\subsection{Cross-directional DC gain matrix}



To determine the cross-directional coupling, we now fix a nominal operating point with constant inlet flows $\mathbf{q}_0:=[\,q_{1,0},\dots,q_{n,0}\,]^\top$ producing steady-state thicknesses $\mathbf{h}_0:=[\,h_{1,0},\dots,h_{n,0}\,]^\top$, so that $q_i(t)=q_{i,0}+\delta q_i(t)$ and $h_j(t)=h_{j,0}+\delta h_j(t)$. After linearising the Navier-Stokes-VOF model around $(\mathbf{h}_0,\mathbf{q}_0)$,  small steady-state deviations $\delta h_j(t)$ depend linearly on small steady-state perturbations $\delta q_i(t)$. Linearising the steady-state map $\mathbf{q}\mapsto h_j$ at $\mathbf{q}=\mathbf{q}_0$ yields the static sensitivity relation
\begin{equation}
\delta h_j
= \sum_{i=1}^n H_{ji}\,\delta q_i,\qquad j=1,\dots,n,
\label{eq:static_sensitivity_relation}
\end{equation}
with exact PDE-level sensitivities
\begin{equation}
H_{ji}
:= \left.\frac{\partial h_j}{\partial q_i}\right|_{\mathbf{q}_0}
\!\!= \frac{1}{A_{\mathrm{sens}}}
\int_{{\mathcal{S}}(x_s,y_j)}\!\!\!\!\!
\left.\frac{\partial \alpha(x_s,y,z,t)}{\partial q_i}\right|_{\mathbf{q}_0}
\!\!dV.
\label{eq:H_sensitivity_exact}
\end{equation}
The scalar $H_{ji}$ quantifies the contribution of the $i$-th inlet stripe to the
steady-state thickness at the $j$-th sensor: it is defined directly from the
PDE-based measurement formulas~\eqref{eq:thickness_field} through the sensitivity of the VOF field $\alpha$ with respect to the inlet
flows. Here $\alpha(x,y,z,t)$ denotes the VOF field of the full
Navier-Stokes-VOF model; for each constant inlet vector $\mathbf{q}$ the PDE
admits a steady solution that depends smoothly on $\mathbf{q}$, and the
derivative $\partial \alpha/\partial q_i$ in
\eqref{eq:H_sensitivity_exact} is taken with respect to this steady-state
dependence and evaluated at $\mathbf{q}=\mathbf{q}_0$. Relation \eqref{eq:static_sensitivity_relation} captures only the steady-state part of
the inlet-to-sensor mapping; the full dynamic response is obtained by combining
this static map with the scalar inlet-to-sensor transfer function $G(s)$
derived in the previous subsection. In what follows we choose the scalar
surrogate $G(s)$ with unit DC gain, $G(0)=1$, so that $\mathbf{H}$ is both the
PDE-level steady-state sensitivity matrix and the DC gain matrix of the
dynamic multiple-input multiple-output (MIMO) model.

Computing the full three-dimensional sensitivities $\partial \alpha/\partial
q_i$ in~\eqref{eq:H_sensitivity_exact} is expensive and problem-specific. In
the thin-film regime, however, the depth-averaged coating equations provide a
simpler description of how cross-directional non-uniformities in the per-width
flow are redistributed before reaching the sensors. Linearising these
thin-film equations leads to an integral representation of the form
\begin{equation}
\delta h(x_s,y,t)\approx \int_0^w K(y,\xi)\,\delta\gamma(\xi,t)\,d\xi,
\label{eq:kernel_representation}
\end{equation}
where $\delta\gamma(\xi,t)$ is the per-width flow perturbation at
cross-directional location $\xi$, and $K(y,\xi)$ is a Green's kernel that
describes how a local perturbation at $\xi$ affects the thickness at $y$ along
the cross direction; see, for example, standard thin-film and coating-flow
treatments such as \citep{Kistler1997,Ruschak1985}. In this view, the exact
sensitivities in~\eqref{eq:H_sensitivity_exact} are approximated, after depth
averaging and restriction to $x=x_s$, by the kernel $K(\cdot,\cdot)$ composed
with appropriate inlet and sensor shapes.

In particular, let $w>0$ be the coating width in the cross
direction, and let $\phi_i(\xi)$ denote the cross-directional distribution of
per-width flow associated with a unit perturbation in the $i$-th inlet. Small
inlet-flow perturbations then generate a flow perturbation profile
\[
\delta\gamma(\xi,t)
= \sum_{i=1}^n \delta q_i(t)\,\phi_i(\xi),\qquad \xi\in[0,w].
\]
Similarly, the finite footprint of the $j$-th sensor can be represented by a
nonnegative aperture function $\psi_j(y)$, supported near $y_j$ and normalised
so that
\[
h_j(t)\approx \int_0^w \psi_j(y)\,h(x_s,y,t)\,dy.
\]
Linearising this expression yields the perturbed sensor output
\[
\delta h_j(t)\approx \int_0^w \psi_j(y)\,\delta h(x_s,y,t)\,dy.
\]

Combining~\eqref{eq:kernel_representation} with the inlet
and sensor shapes, we obtain
\begin{align*}
\delta h_j(t)
&\approx \int_0^w \psi_j(y)\left[\int_0^w K(y,\xi)\,\delta\gamma(\xi,t)\,d\xi\right]dy\\
&= \int_0^w \int_0^w \psi_j(y)\,K(y,\xi)\left[\sum_{i=1}^n \delta q_i(t)\,\phi_i(\xi)\right] d\xi\,dy\\
&= \sum_{i=1}^n \delta q_i(t)\left[\int_0^w \int_0^w \psi_j(y)\,K(y,\xi)\,\phi_i(\xi)\,d\xi\,dy\right].
\end{align*}
The bracketed term defines the cross-directional DC gain from inlet $i$ to
sensor $j$,
\begin{equation}
H_{ji}\approx \int_0^w \int_0^w \psi_j(y)\,K(y,\xi)\,\phi_i(\xi)\,d\xi\,dy,
\label{eq:H_from_kernel_general}
\end{equation}
which can be viewed as a thin-film-based approximation of the exact
sensitivities in~\eqref{eq:H_sensitivity_exact}. Each $H_{ji}$ aggregates
three effects: the actuation pattern across the width ($\phi_i$), the
spreading of thickness perturbations by the coating manifold ($K$), and the
spatial averaging performed by the sensor ($\psi_j$). Constructing a
PDE-motivated parametric family for the matrix $\mathbf{H}=[H_{ji}]\in\mathbb R^{n\times n}$ is
therefore a central step of the surrogate.

Collecting the perturbations into vectors $\delta\mathbf{h}(t)$ and $\delta\mathbf{q}(t)$, the cross-directional mapping can be written as
\begin{equation}
\delta\mathbf{h}(t)\approx \mathbf{H}\,\delta\mathbf{q}(t).
\label{eq:delta_h_vs_delta_q}
\end{equation}
Because all channels share the same convective-relaxation dynamics encoded in the scalar transfer function $G(s)$ in~\eqref{eq:G_def} for a representative inlet and its downstream thickness, the full MIMO plant factors as a common scalar dynamic multiplying this static influence matrix:
\begin{equation}
\mathbf{Y}(s) = G(s)\,\mathbf{H}\,\mathbf{Q}(s),
\label{eq:mimo_factor}
\end{equation}
where $\mathbf{Q}(s)$ and $\mathbf{Y}(s)$ are the Laplace transforms of the
small-signal inlet-flow and thickness vectors, respectively.

\section{Parameter Identification}
\label{sec:pde-informed-H}

Building on the \texttt{interFoam} model in Section~\ref{sec:pde-model} and the low-order structure in Section~\ref{sec:pde-to-ode}, we now show how the scalar surrogate $G(s)$ and the cross-directional DC gain matrix $\mathbf H$ are tied back to the PDE and how their coefficients are chosen and validated for the present configuration.

We consider $n=5$ inlet stripes in the cross direction $y$ centred at $y_i\in\{15,45,75,105,135\}$ \si{\milli\meter}. Each stripe is a \texttt{flowRateInletVelocity} patch of width $w_s=\SI{30}{\milli\meter}$, channel height $h_{\mathrm{ch}}=\SI{4}{\milli\meter}$, and area $A_{\mathrm{stripe}}=w_s\times h_{\mathrm{ch}}=\SI{120}{\milli\meter^2}$. The boundary condition prescribes a volumetric flow $q_i(t)$ through each stripe, applied uniformly over $A_{\mathrm{stripe}}$. Downstream, five virtual thickness sensors are located at machine position $x_s = \SI{30}{\milli\meter}$ and the same cross-directional positions as the inlets. Each sensor performs a \texttt{volIntegrate} of the liquid volume fraction $\alpha$ over a right cylinder with a radius of $r_c=\SI{1}{\milli\meter}$ and a height of \SI{1}{\milli\meter}, so that the cylinder area is $A_{\mathrm{sens}} = \pi r_c^2$. The local thickness is defined as the cylinder volume divided by $A_{\mathrm{sens}}$, yielding a scalar time series $h_j(t)$ at each sensor.

The \texttt{interFoam} case uses a Carreau-Yasuda liquid and air as the two phases, with liquid density $\rho_{\ell}=1.20\times 10^{3}\si{\kilo\gram\per\meter^3}$, air density $\rho_{a}=1.0\ \mathrm{kg/m^3}$, air viscosity $\mu_a = 1.5\times 10^{-5}\ \mathrm{Pa\cdot s}$, surface tension $\sigma = 3.5\times 10^{-2}\ \mathrm{N/m}$, and a dynamic contact-angle condition with advancing and receding angles $\theta_A = 40^\circ$ and $\theta_R = 30^\circ$. The Carreau-Yasuda viscosity law for the liquid is characterised by $\mu_0 = 10\ \mathrm{Pa\cdot s}$, $\mu_{\infty} = 0.1\ \mathrm{Pa\cdot s}$, $\lambda = 0.1\ \mathrm{s}$, $m = 0.6$, $a = 2$. At the operating point the substrate speed is $U_0 = 0.333\ \mathrm{m/s}$, the steady film thickness at the sensor station is $h_0 \approx 8.7\times 10^{-5}\ \mathrm{m}$, the coating width is $w = 0.15\ \mathrm{m}$, and the baseline volumetric flows are $q_{i,0}=1.0\times 10^{-6}\ \mathrm{m^3/s}$ for $i=1,\dots,5$.

Starting from a nominal state $(U_0,h_0,\mathbf q_0)$, we excite each of the five inlet flows with an independent $10\%$ pseudo-random binary sequence (PRBS) around $q_{i,0}$, with a bit length of \SI{0.01}{\second} and a total duration of \SI{2}{\second}. The resulting inlet-flow signals $q_i(t)$ and sensor thicknesses $h_j(t)$ are logged at a sampling time $T_s=\SI{0.01}{\second}$. All subsequent identification uses this single \SI{2}{\second}-long, open-loop PRBS dataset.

For the depth-averaged analysis in Section~\ref{sec:pde-to-ode} we approximate the Carreau-Yasuda rheology by a Newtonian liquid with an effective constant viscosity at $(U_0,h_0,\mathbf q_0)$, in order to obtain a simple convective-relaxation PDE with constant coefficients. Along the machine direction $x$, the PDE and the geometry fix the transport time between the inlets and the sensor locations. With sensors at $x_s$ and substrate speed $U_0=0.333\ \mathrm{m/s}$, the thin-film convective model \eqref{eq:convective_relaxation_PDE} predicts
\[
L_{\mathrm{PDE}} = \frac{x_s}{U_0} = \frac{0.03}{0.333} \approx 0.09009\ \mathrm{s}.
\]
To check this PDE prediction against the CFD data, we first fit a dead-time second-order-with-zero model to a representative inlet-sensor pair (inlet 3 to sensor 3) from the same PRBS dataset, with $L$, $b_0$, $b_1$, $c_0$, and $c_1$ treated as free parameters. This SISO parametric fit yields an identified delay
\[
L_{\mathrm{id}}\approx 0.090\ \mathrm{s},
\]
which agrees with $L_{\mathrm{PDE}}$ to within less than $0.1\%$. We therefore fix $L$ as $L= \SI{0.09}{\second}$ and interpret $L$ as a transport delay set directly by the convective part of the PDE; numerically it is consistent with the identified value $L_{\mathrm{id}}$.

The remaining inlet-to-sensor dynamics arise from storage and relaxation in the coating-supply system described by \eqref{eq:convective_relaxation_PDE}-\eqref{eq:two_state_model}, which lead to the delay-free SISO transfer function $G_{0}(s)$ in \eqref{eq:G_relax_rational}. Here, the numerator coefficient $b_0$ is constrained to equal the stiffness coefficient $c_0$, so $G_0(0)=1$. In practice, the parameters $(b_0,b_1,c_0,c_1)$ are then refined by a time-domain least-squares fit on the same SISO channel (inlet~3 to sensor~3). We work with deviation variables around the nominal operating point, fix the delay to $L_{\mathrm{PDE}}$, and discretise the delay-free second-order model by a zero-order hold at the CFD sampling time $T_s=\SI{0.01}{\second}$. The resulting discrete-time model is simulated over the PRBS dataset, and the coefficients are chosen to minimise the sum of squared errors between the simulated thickness and the \texttt{interFoam} thickness trace over the full PRBS dataset. Once this continuous-time calibration is completed, $G_0(s)$ is assumed to be fixed. In the MIMO identification below, the same zero-order-hold discretisation is used with $T_s=0.01\,$s when constructing discrete-time regressors, so that the SISO fit and the MIMO regression are consistent. This SISO regression yields $
b_1 \approx 5.28\times 10^{-4}$, $c_1 \approx 1.97\times 10^{2}$, and $c_0 = b_0 \approx 1.87\times 10^{4}$, so that the associated real zero is located at $s_z\approx -b_0/b_1\approx -3.5\times 10^{7}\ \mathrm{s^{-1}}$. This zero is several orders of magnitude faster than the inverse sampling time and is therefore only weakly excited by the $0.01\,$s PRBS signals; over the frequency range of interest the response is dominated by the second-order convective-relaxation dynamics. Motivated by this observation, in the surrogate used below we set $b_1=0$.

Combining this delay-free model with the transport delay $L$ gives the scalar transfer function $G(s)\ =\ e^{-L s}G_0(s)$, which is shared
by all inlet channels. In the MIMO configuration we apply this scalar dynamic
channel-wise to the inlet deviations and then mix the resulting filtered
signals through a static cross-directional gain. In the Laplace domain this
corresponds to
\begin{equation}
\delta\mathbf h(s)\ \approx\ G(s)\,\widehat H\,\delta\mathbf q(s),
\end{equation}
where $G(s)$ acts identically on each inlet channel. Equivalently, in the time
domain we first form the filtered inlet signals
\[
\mathbf r(t)=(G*\delta\mathbf q)(t),
\]
and then apply the static mixing
\[
\delta\mathbf h(t)\approx \widehat H\,\mathbf r(t).
\]
Here, $\widehat{H}$ is the identified version of the cross-directional DC gain matrix $\mathbf{H}$.

With the scalar dynamics $G(s)$ fixed by the SISO fit, the cross-directional DC gain matrix $\widehat H\in\mathbb{R}^{5\times 5}$ is identified from the full five-input five-output PRBS dataset. We form the regressor signals
\[
\mathbf r(t) = \big[r_1(t),\dots,r_5(t)\big]^\top,
\qquad
r_i(t) = (G * \delta q_i)(t),
\]
where $(G * \delta q_i)(t)$ denotes the output of the fixed SISO LTI system
with transfer function $G(s)$ when driven by the $i$-th inlet-flow deviation
$\delta q_i(t)$, including the delay $L$. In the discrete-time implementation
used here, the delay $L$ is represented by a $9$-sample shift and the
delay-free part of $G$ is implemented using a zero-order-hold discretisation, so that each $r_i(k)$ is obtained by filtering the delayed input $\delta q_i(k)$ through the discrete-time counterpart of $G$. Stacking the samples over the full $0$-$2\,$s horizon then yields a linear regression problem of the form
\[
\delta\mathbf h(k) \approx \widehat H\,\mathbf r(k),\qquad k=1,\dots,N,
\]
which is solved in a least-squares sense, row by row, to obtain the entries of $\widehat H$.

Carrying out this procedure on the \texttt{interFoam} logs gives
\[
\widehat H
\approx
\begin{bmatrix}
49.38 &  1.03 &  7.41 & 16.54 & 14.53\\
 9.45 & 43.59 & 11.75 & 10.76 & 13.14\\
 1.39 &  5.58 & 52.63 & 13.37 & 15.49\\
 3.34 & 13.50 &  6.08 & 53.15 & 12.61\\
 0.16 &  4.64 & 11.19 & 18.51 & 54.84
\end{bmatrix}
\ \left[\frac{\mathrm{m}}{\mathrm{m^3/s}}\right].
\]
The diagonal entries are all of order $50\,\mathrm{m}/(\mathrm{m^3/s})$, consistent with the dominant self-coupling of each inlet-sensor pair, while nearest-neighbour and more distant couplings lie between $O(10)$ and $O(1)$ and are not symmetric across the diagonal. These long-range and asymmetric terms reflect the details of the manifold and plenum geometry as resolved by the full \texttt{interFoam} model.

To compare this data-driven cross-directional map with a PDE model, we now return to the depth-averaged kernel description of Section~\ref{sec:pde-to-ode}. In that reduced description, the linearised cross-directional mechanics are summarised by a kernel $K(y,\xi)$ that maps per-width flow perturbations $\delta\gamma(\xi,t)$ to thickness perturbations $\delta h(x_s,y,t)$ at the sensor station as in \eqref{eq:kernel_representation}, leading to
\begin{equation}
H_{ji}\approx\int_0^w\!\!\int_0^w \psi_j(y)\,K(y,\xi)\,\phi_i(\xi)\,d\xi\,dy,
\label{eq:H_from_kernel_general_again_short}
\end{equation}
where $\phi_i$ describes the cross-directional actuation pattern of inlet $i$, $\psi_j$ the sensor footprint, and $K$ the redistribution by the coating manifold.

On the actuation side, the stripe boundary condition introduced above can be written in terms of a per-width flow $\gamma_i(y,t)$ [m$^2$/s], obtained by dividing the total inlet flow $q_i(t)$ by the stripe width and approximated as uniformly distributed over each stripe:
\begin{align*}
\gamma_i(y,t)&\approx
\begin{cases}
\dfrac{q_i(t)}{w_s}, & y\in\big[y_i-\tfrac{w_s}{2},\,y_i+\tfrac{w_s}{2}\big],\\[0.5em]
0, & \text{otherwise},
\end{cases}
\end{align*}
where $y_i\in\{15,45,75,105,135\}\ \mathrm{mm}$. In the notation of Section~\ref{sec:pde-to-ode} this corresponds to a stripe shape
\[
\phi_i(\xi)=
\begin{cases}
\dfrac{1}{w_s}, & \xi\in\big[y_i-\tfrac{w_s}{2},\,y_i+\tfrac{w_s}{2}\big],\\[0.5em]
0, & \text{otherwise},
\end{cases}
\]
so that $\delta\gamma(y,t)=\sum_{i=1}^5 \delta q_i(t)\,\phi_i(y)$.

On the sensing side, the volume-integrating cylinders defined above correspond, in the depth-averaged model, to point evaluations at $y=y_j$ because $r_c\ll w_s$. We therefore approximate the sensor footprint by $\psi_j(y)\approx \delta(y-y_j)$, which reduces \eqref{eq:H_from_kernel_general_again_short} to
\begin{equation}
H_{ji}\approx\int_0^w K(y_j,\xi)\,\phi_i(\xi)\,d\xi.
\label{eq:H_point_sensor_short}
\end{equation}

Under manifold-dominated cross-directional equalisation the cross-directional response is diffusion-like, symmetric, and short-ranged, so we model the Green's kernel by a symmetric Gaussian profile
\[
K(y,\xi)=\kappa\,\exp\!\Big(-\frac{(y-\xi)^2}{2\ell^2}\Big),
\]
where $\kappa>0$ sets the overall scale and $\ell>0$ is an effective cross-directional spread. Substituting this kernel and the stripe shape $\phi_i$ into \eqref{eq:H_point_sensor_short} gives
\[
H_{ji}(\kappa,\ell)
= \frac{\kappa}{w_s}
\int_{y_i-\frac{w_s}{2}}^{y_i+\frac{w_s}{2}}
\exp\!\Big(-\frac{(y_j-\xi)^2}{2\ell^2}\Big)\,d\xi,
\]
which evaluates as
\begin{align*}
&H_{ji}(\kappa,\ell)\\
&=\frac{\kappa\,\sqrt{\pi/2}\,\ell}{w_s}
\left[
\operatorname{erf}\!\Big(\frac{y_i+\frac{w_s}{2}-y_j}{\sqrt{2}\,\ell}\Big)
-\operatorname{erf}\!\Big(\frac{y_i-\frac{w_s}{2}-y_j}{\sqrt{2}\,\ell}\Big)
\right],
\end{align*}
with $\operatorname{erf}(\eta)=\frac{2}{\sqrt{\pi}}\int_0^{\eta} e^{-t^2}dt$. Collecting these entries into a matrix, we write
\[
H(\kappa,\ell)\ :=\ \big[\,H_{ji}(\kappa,\ell)\,\big]_{j,i=1}^5.
\]
This defines a two-parameter family $H(\kappa,\ell)$ of symmetric, numerically banded cross-directional maps.

To compare this PDE-motivated family with the data-identified map $\widehat H$, we again separate shape and scale. For each $\ell>0$ we fix $\kappa=1$ and define
\[
H_0(\ell)\ :=\ H(1,\ell),
\]
so that $H(\kappa,\ell)=\kappa\,H_0(\ell)$. We then choose $\kappa^\star$ and $\ell^\star$ so that $H(\kappa^\star,\ell^\star)$ best matches $\widehat H$ in Frobenius norm. Writing
\[
\langle A,B\rangle_F = \mathrm{trace}(A^\top B),\qquad
\|A\|_F = \sqrt{\langle A,A\rangle_F},
\]
we set, for each $\ell>0$,
\[
\kappa^\star(\ell)=\frac{\langle \widehat H,\ H_0(\ell)\rangle_F}{\|H_0(\ell)\|_F^2},
\ell^\star\in\arg\min_{\ell>0}\ \big\|\,\widehat H-\kappa^\star(\ell)\,H_0(\ell)\,\big\|_F.
\]
For the present data this yields $\ell^\star\simeq 1.40\times 10^{-2}\ \mathrm{m}$ and $\kappa^\star\simeq 60.6$. The resulting PDE-motivated, calibrated map $H_{\mathrm{PDE}}=\kappa^\star\,H_0(\ell^\star)$ (five inlets and sensors at $y=\{15,45,75,105,135\}\ \mathrm{mm}$, $w_s=30\ \mathrm{mm}$) can be written compactly, with entries rounded to three significant figures, as
\[
H_{\mathrm{PDE}}
\approx
\begin{bmatrix}
50.7 &  9.94 & 0.044 & 0     & 0\\
 9.94 & 50.7 & 9.94  & 0.044 & 0\\
0.044 & 9.94 & 50.7  & 9.94  & 0.044\\
0     & 0.044 & 9.94 & 50.7  & 9.94\\
0     & 0     & 0.044 & 9.94 & 50.7
\end{bmatrix}
\ \left[\frac{\mathrm{m}}{\mathrm{m^3/s}}\right].
\]

By construction, the kernel-based map $H_{\mathrm{PDE}}$ has nonnegative entries and, after rounding very small entries to zero, is symmetric and effectively banded: both the actuator footprints $\phi_i$ and the sensor apertures $\psi_j$ are nonnegative, and the Gaussian kernel $K(y,\xi)$ is strictly positive but short-ranged. In contrast, the data-identified cross-directional matrix $\widehat H$ exhibits appreciable asymmetries and non-negligible long-range couplings (e.g.\ from inlet~1 to sensor~4 and~5). These effects reflect out-of-model features such as global mass-conservation constraints and the detailed three-dimensional manifold geometry. They contribute to a relative Frobenius norm error
\[
\frac{\big\|\,\widehat H-H_{\mathrm{PDE}}\,\big\|_F}{\|\widehat H\|_F}
\approx 0.33,
\]
and highlight the limits of a purely local, diffusion-like kernel as a cross-directional model. At the same time, the Gaussian family suggested by the thin-film PDE captures the dominant diagonal structure and the short-range nearest-neighbour coupling, while the scalar dynamics $G(s)$ represent a PDE-informed, data-calibrated surrogate for the shared convective-relaxation behaviour along the machine direction.

\section{Validation and Feedback Control}
\label{sec:time-domain-validation}

\begin{figure}[t]
  \centering
  \includegraphics[width=\columnwidth]{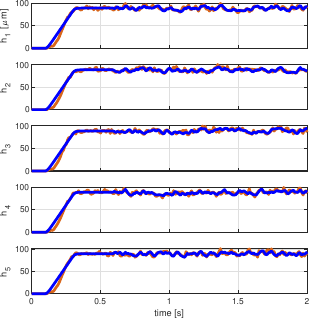}
  \caption{Time-domain comparison between the \texttt{interFoam} film thickness $h_j(t)$ (orange) and the identified surrogate response (blue) at the five sensors under PRBS inlet-flow perturbations.}
  \label{fig:time-domain-mimo}
\end{figure}

We first validate the identified MIMO surrogate of Section~\ref{sec:pde-informed-H}, whose small-signal Laplace-domain mapping is
\[
\delta\mathbf h(s)\ \approx\ G(s)\,\widehat H\,\delta\mathbf q(s),
\]
against the $0\text{-}2\,\mathrm{s}$ open-loop \texttt{interFoam} logs by driving it with the measured inlet-flow deviations $\delta\mathbf q(t)=\mathbf q(t)-\mathbf q_0$ and comparing the predicted thickness trajectories with the CFD outputs at all five sensors. At $t=0$ the nominal sensor thickness vector is
\[
\mathbf h_0 := \mathbf h(0)\ \approx\ 
\big[\,90.51,\ 87.16,\ 84.02,\ 88.30,\ 90.85\,\big]\ \mu\mathrm{m},
\]
and we denote its components by $h_{j,0}$, $j=1,\dots,5$. The corresponding mean thickness is
\[
\bar h_0\ \approx\ 88.17\,\mu\mathrm{m}.
\]
Over the full record the per-sensor root-mean-square errors between surrogate and CFD lie in the range $3.7\text{-}4.0\,\mu\mathrm{m}$ (about $4\text{-}4.5\%$ of the nominal thickness), with coefficients of determination $R^2_j\approx 0.98$ for all five channels. The time-domain comparison in Fig.~\ref{fig:time-domain-mimo} shows that the identified model reproduces the transport delay, dominant relaxation, and steady-state gains of the CFD dynamics to within a few percent.

On this surrogate we consider a simple P controller in deviation coordinates,
\[
\delta\mathbf q(t)
\ =\
K_P\big(\delta\mathbf h_{\mathrm{ref}}-\delta\mathbf h(t)\big),
\]
where $\delta\mathbf q(t)=\mathbf q(t)-\mathbf q_0$ and
$\delta\mathbf h(t)=\mathbf h(t)-\mathbf h_0$ denote deviations from the nominal operating point $(\mathbf q_0,\mathbf h_0)$ of Section~\ref{sec:pde-informed-H}, and $\delta\mathbf h_{\mathrm{ref}}=\mathbf h_{\mathrm{ref}}-\mathbf h_0$ is the reference in deviation coordinates. Because $G(0)=1$, the steady-state relation between inlet-flow deviations and thickness deviations is
\[
\delta\mathbf h_\infty = \widehat H\,\delta\mathbf q_\infty,
\]
where $\delta\mathbf h_\infty := \lim_{t\to\infty}\delta\mathbf h(t)$ and $\delta\mathbf q_\infty := \lim_{t\to\infty}\delta\mathbf q(t)$ denote the steady-state deviation vectors. We choose
\[
K_P\ :=\ \beta\,\widehat H^{-1},\qquad \beta = 0.1,
\]
so that the nominal closed-loop DC map satisfies
\[
\delta\mathbf h_{\infty}
= \frac{\beta}{1+\beta}\,
\delta\mathbf h_{\mathrm{ref}},
\]
i.e. a tracking fraction of $\beta/(1+\beta)\approx 0.091$ on all modes. For the present data this gives the P gain matrix
\[
K_P\ \approx\
\begin{bmatrix}
 0.0020 &  0.0002 & -0.0002 & -0.0005 & -0.0004\\
-0.0004 &  0.0024 & -0.0004 & -0.0001 & -0.0003\\
 0.0000 & -0.0001 &  0.0021 & -0.0003 & -0.0005\\
 0.0000 & -0.0006 &  0.0000 &  0.0021 & -0.0003\\
 0.0000 &  0.0000 & -0.0004 & -0.0006 &  0.0021
\end{bmatrix}.
\]

\begin{figure}[t]
  \centering
  \includegraphics[width=\columnwidth]{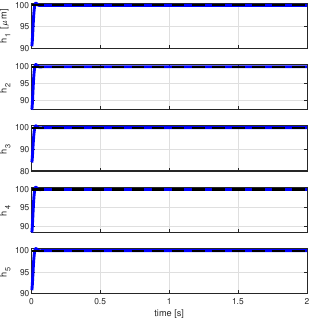}
  \caption{Closed-loop absolute thickness responses $h_j(t)$ at the five sensors under the P controller.}
  \label{fig:closed-loop-mimo}
\end{figure}

To drive the absolute thickness at all sensors to a common target $h_{\mathrm{tar}} = 100\,\mu\mathrm{m}$, we choose the deviation reference so that the steady-state relation
$\delta\mathbf h_{\infty} = \tfrac{\beta}{1+\beta}\,\delta\mathbf h_{\mathrm{ref}}$
implies, for each sensor,
\[
h_{j,\infty}
= h_{j,0} + \delta h_{j,\infty}
= h_{\mathrm{tar}},\qquad j=1,\dots,5,
\]
where $h_{j,\infty}$ and $\delta h_{j,\infty}$ denote the $j$-th components of the steady-state absolute thickness vector $\mathbf h_\infty := \lim_{t\to\infty}\mathbf h(t)$ and the deviation vector $\delta\mathbf h_\infty$, respectively. Writing $\delta\mathbf h_{\mathrm{ref}} = [\,\delta h_{\mathrm{ref},1},\dots,\delta h_{\mathrm{ref},5}\,]^\top$, with $\delta h_{\mathrm{ref},j}$ the $j$-th component, this yields, componentwise,
\[
\delta h_{\mathrm{ref},j}
= \frac{h_{\mathrm{tar}}-h_{j,0}}{\beta/(1+\beta)},\qquad j=1,\dots,5,
\]
so that, for the nominal thickness vector $\mathbf h_0$ above, the required reference in deviation coordinates is
\[
\delta\mathbf h_{\mathrm{ref}}
\approx
\big[\,104.42,\ 141.19,\ 175.83,\ 128.65,\ 100.64\,\big]\ \mu\mathrm{m}.
\]
Under this P controller with $\beta=0.1$, the surrogate exhibits well-damped but relatively slow closed-loop responses in absolute thickness: each $h_j(t)$ starts from its nominal value $h_{j,0}$ and converges monotonically to the common $100\,\mu\mathrm{m}$ target, with the cross-directional profile remaining nearly uniform; see Fig.~\ref{fig:closed-loop-mimo} for the absolute thickness trajectories.

\section{Conclusion}

Starting from a two-phase incompressible Navier-Stokes-VOF model of a slot-die coating slice, we have built a compact surrogate that separates convective transport along the machine direction from cross-directional coupling through a static DC gain matrix, with all parameters calibrated directly to CFD data. On top of this PDE-informed plant we designed a simple proportional controller, expressed in deviation coordinates but tuned to drive the absolute film thickness at all sensors toward a common target while maintaining a nearly uniform cross-directional profile. The results show that a modest amount of identification, guided by the governing equations and the actual slot-die geometry, is sufficient to obtain a transparent, product-centric feedback design in simulation.

The structure of the surrogate is intended to extend beyond the initial configuration studied here. Different slot-die geometries, for example alternative manifolds, plenum layouts, or die-lip shapes, will change the scalar convective-relaxation dynamics and the cross-directional DC gain matrix, but the same PDE-informed pipeline can be used to re-identify a shared dynamic factor and a geometry-dependent influence matrix from CFD or experimental data. The comparison between the symmetric, short-ranged kernel-based map and the fully identified cross-directional gain already reveals how edge effects, global mass conservation, and manifold non-uniformities manifest as asymmetries and long-range couplings, providing a systematic way to diagnose and eventually mitigate edge behavior. Finally, many practically relevant lines are underactuated or overactuated in the cross direction, with fewer or more inlet zones than sensor locations; in such cases the same framework naturally leads to rectangular gain matrices and controller designs based on regularized inverses that trade off overall loading, profile uniformity, and actuation effort. Exploring these underactuated layouts, more complex die geometries, and CFD in-the-loop and experimental implementations are natural next steps toward deploying PDE-informed surrogate control on industrial electrode coating lines. The cross-directional model developed in this paper represents a first step towards those advanced control methods.

\end{document}